\documentclass[lettersize,journal]{IEEEtran}
\usepackage[utf8]{inputenc}

\usepackage{cite}
\usepackage{amsmath,amssymb,amsfonts}
\usepackage{algorithmic}
\usepackage{array}
\usepackage{textcomp}
\usepackage{stfloats}
\usepackage{url}
\usepackage{verbatim}
\usepackage{graphicx}
\usepackage{balance}

\usepackage[caption=false,font=normalsize,labelfont=sf,textfont=sf]{subfig}

\usepackage{xcolor}
\usepackage{booktabs}
\usepackage[hidelinks]{hyperref}
\usepackage{orcidlink}
\usepackage{algorithm}
\usepackage{listings}
\usepackage{multirow}
\usepackage{svg}

\definecolor{lightgreen}{rgb}{0.95, 1.0, 0.95}
\definecolor{codeblue}{rgb}{0.1, 0.1, 0.8}
\definecolor{codegreen}{rgb}{0, 0.5, 0}
\definecolor{codegray}{rgb}{0.5, 0.5, 0.5}

\def\BibTeX{{\rm B\kern-.05em{\sc i\kern-.025em b}\kern-.08em
    T\kern-.1667em\lower.7ex\hbox{E}\kern-.125emX}}

\usepackage[automake,toc,abbreviations]{glossaries-extra}
\newabbreviation{dss}{DSS}{Dynamic Spectrum Sharing}
\newabbreviation{dsm}{DSM}{Dynamic Spectrum Management}
\newabbreviation{dsa}{DSA}{Dynamic Spectrum Access}
\newabbreviation{cr}{CR}{Cognitive Radio}
\newabbreviation{tvws}{TVWS}{TV White Space}
\newabbreviation{cbrs}{CBRS}{Citizen Broadband Radio Service}
\newabbreviation{hlf}{HLF}{Hyperledger Fabric}
\newabbreviation{sdr}{SDR}{Software Defined Radio}
\newabbreviation{pu}{PU}{Primary User}
\newabbreviation{su}{SU}{Secondary User}
\newabbreviation{fcc}{FCC}{Federal Communications Commission}
\makeglossaries

\begin{document}

\title{BLAST: Blockchain-based LLM-powered Agentic Spectrum Trading}

\author{
Anas~Abognah\textsuperscript{\orcidlink{0009-0000-9944-7627}},~\IEEEmembership{Student Member,~IEEE,}
        and~Otman~Basir\textsuperscript{\orcidlink{0000-0002-6454-0538}},~\IEEEmembership{Member,~IEEE}% <-this % stops a space

\thanks{A. Abognah and O. Basir are with the Department of Electrical and Computer Engineering, University of Waterloo, Waterloo, ON N2L 3G1, Canada (e-mail: aabognah@uwaterloo.ca; obasir@uwaterloo.ca).}%

\thanks{This work has been submitted to the IEEE for possible publication. Copyright may be transferred without notice, after which this version may no longer be accessible.}
}

\markboth{Submitted to IEEE TRANSACTIONS on COGNITIVE COMMUNICATIONS AND NETWORKING, VOL. xx, NO. x, March 2026}%
{BLAST: Blockchain-based LLM-powered Agentic Spectrum Trading}

%remove when publishinh)
\IEEEpubid{This work has been submitted to the IEEE for possible publication.}

\maketitle

\begin{abstract}
The management of radio frequency spectrum is undergoing a paradigm shift from static, centralized command-and-control models to dynamic, market-driven approaches. However, the realization of \gls{dsm} has been hindered by the lack of an automated, trustworthy, and intelligent coordination infrastructure that can operate without a central authority while preserving participant privacy. In this paper, we introduce \textit{BLAST} (Blockchain-based LLM-powered Agentic Spectrum Trading), a comprehensive framework that integrates Large Language Model (LLM) Agents with a permissioned blockchain infrastructure to create a fully autonomous, private, and secure spectrum trading ecosystem. We propose a novel agent architecture that implements the Cognitive Radio cycle through a sequential decision pipeline (perceive, plan, act) enabling agents to reason strategically about economic value and market dynamics. We evaluate the framework through three distinct market mechanisms: Direct Sale, First-Price Sealed-Bid, and Second-Price (Vickrey) Sealed-Bid auctions. Experimental results demonstrate that the Second-Price (Vickrey) auction is the optimal choice for maximizing social welfare and allocative efficiency, capturing up to 71\% of the theoretical surplus by incentivizing truthful bidding. We also compare the proposed model against a baseline non-LLM heuristic agentic model and show that utilizing LLM agents yields significant improvements in market competition, reduced wealth and asset concentration, and increased system welfare. Furthermore, we validate the system's privacy preservation, confirming that sensitive bid values remain isolated in private data collections while only cryptographic hashes are committed to the public ledger.
\end{abstract}


\section{Introduction}

The management of radio frequency spectrum is currently undergoing a fundamental transformation, moving away from conventional static, centralized command-and-control regimes toward dynamic, market-driven mechanisms \cite{zhao2007survey, perera2024survey}. This paradigm shift is driven by the increasing scarcity and under-utilization of spectrum resources, which impose a significant impediment to the growth of wireless networks and users in the forthcoming sixth generation (6G) era and beyond \cite{wang2025blockchainenabled, javaid2025agi, perera2024survey}. The realization of genuinely autonomous \gls{dsm} requires an intelligent, trustworthy, and decentralized infrastructure that can efficiently coordinate resource allocation \cite{perera2024survey, zheng2020smartcontract}.

The initial technological solution proposed to address the artificial scarcity created by static allocation was the \gls{cr} \cite{mitola1999cognitive}. \glspl{cr}, defined as intelligent devices capable of sensing and adapting to the radio environment, enabled the concept of \gls{dsa} \cite{haykin2005cognitive}. However, \gls{cr} alone proved insufficient to provide a fully dynamic ecosystem that could guarantee protection for spectrum owners and reliably manage complex multi-user interactions \cite{FCC_CR}. Consequently, many spectrum regulators pursued a centralized spectrum management model, such as the database-driven systems for \gls{tvws} and \gls{cbrs} \cite{abognah_tvws_libya,cbrs_intro_paper}.

These centralized approaches were ultimately hindered by several critical drawbacks, failing to deliver the full vision of robust and market-driven \gls{dsm} \cite{yang2022decentralized_spectrum_access_bc}:
\begin{enumerate}
    \item \textit{Single Point of Failure and Scalability:} A centralized database or broker is susceptible to a single point of failure and lacks the necessary scalability to support the exponential growth of distributed devices in future networks (e.g., IoT and 6G). Furthermore, traditional centralized network management models struggle to meet the requirements of highly dynamic environments. 
    \item \textit{Trust and Transparency Issues:} These models require participating entities (Primary and Secondary Users) to entrust sensitive operational data to an untrusted spectrum broker. This arrangement creates severe risks of privacy disclosure, data tampering, and a lack of transparency regarding transaction records, as the regulator must verify the database state without intrinsic user-verifiable mechanisms.
    \item \textit{Lack of Incentive Compatibility:} Centralized systems typically do not intrinsically implement a spectrum trading system, providing no native incentive mechanism for licensed users (PUs) to share their spectrum.
\end{enumerate}

% Somewhere in the second column of the first page (remove for publishing in IEEE):
\IEEEpubidadjcol

The realization of effective \gls{dsm}, therefore, requires a concerted shift toward decentralization that distributes trust, control, and decision-making authority while preserving privacy and providing incentive mechanisms. These requirements are addressed through Blockchain technology, which offers an immutable and distributed ledger suitable for securely recording spectrum access rights and transaction history \cite{abognah2022distributed}. In addition, the use of Smart Contracts (SCs) enables the autonomous execution of auction protocols, enhancing data verifiability and fairness in bidding platforms and incentivizing spectrum owners to participate in profit maximizing trades.

Multiple blockchain platforms have been proposed in the literature for various applications and use cases. For enterprise adoption, permissioned architectures such as Hyperledger Fabric are favored for their ability to ensure transaction privacy and commercial confidentiality, often utilizing sophisticated mechanisms like private data collections to hide sensitive bid values \cite{hyperledgerfabric}. 

Alongside blockchain, the realization of genuinely autonomous \gls{dsm} requires an intelligent, trustworthy, and decentralized infrastructure that can efficiently coordinate resource allocation. In recent years much research in \gls{dsa} leveraged sophisticated Artificial Intelligence (AI) techniques, particularly Deep Reinforcement Learning (DRL) and Multi-Agent Reinforcement Learning (MARL), for numerical optimization and low-level control functions such as spectrum sensing and interference avoidance \cite{jiang2019multi-agent-dsa, gao2021multi-agent-coop-ss, ukpong2025deep, vangaru2024multiagent, chen2025radiollm}. However, these conventional AI approaches often lack the high-level, human-like reasoning, strategic interpretation, and policy comprehension essential for participating in complex economic markets \cite{lee2024llmempowered, karim2025aiagents}. Furthermore, most MARL systems proposed in the literature employ centralized optimization methods which are functionally constrained by network size, frequently generating hallucinatory or incoherent outputs when forced to process large amounts of global state information, thus degrading optimization performance \cite{lee2024malo}. Decentralization enhances scalability by allowing agents to focus on local solutions, which reduces response length and improves reasoning capabilities.

The imperative for sophisticated decision-making for decentralized \gls{dsm} systems coincides with the emergence of Large Language Models (LLMs) and their deployment as autonomous, goal-oriented computational entities, or LLM Agents \cite{karim2025aiagents, qu2025llmenabled, lee2024malo}. LLM agents, equipped with advanced planning and reasoning capabilities, have shown immense potential in revolutionizing wireless systems, ranging from accelerating complex policy-making and regulatory workflows, such as data extraction and consultation synthesis \cite{rutagemwa2024empowering}, to enabling end-to-end intelligent resource management and orchestration in next-generation networks \cite{lee2024malo, javaid2025agi}. Furthermore, pioneering frameworks have demonstrated the feasibility of integrating LLMs directly into physical layer signal processing tasks like classification, effectively bridging the modality gap between language and radio signals \cite{chen2025radiollm, zhou2025spectrumfm}.

In this paper, we introduce \textit{\textit{BLAST}} (Blockchain-based LLM-powered Agentic Spectrum Trading), a comprehensive framework that integrates Large Language Model (LLM) Agents with a permissioned blockchain infrastructure to create a fully autonomous, private, and secure spectrum trading ecosystem. This system is designed to reconcile the advanced strategic intelligence of LLM agents with the structural integrity of a decentralized mechanism. We propose a novel agent architecture that implements the foundational Cognitive Radio (CR) cycle through a rigorous sequential decision pipeline. This pipeline, comprising stages such as historical analysis, market hypothesis formation, strategy optimization, and action execution, enables agents to reason strategically and autonomously about economic value and dynamic market conditions. 

Furthermore, the framework is implemented using a enterprise-grade Hyperledger Fabric network to enforce critical privacy guarantees, ensuring commercial confidentiality through the use of private data collections and a commit-reveal protocol. By leveraging the incentive compatibility of the Second-Price (Vickrey) Sealed-Bid Auction \cite{vickrey1961counterspeculation}, the proposed framework provides a provably truthful mechanism that guarantees the maximization of social welfare regardless of the agents' inherent bounded rationality.

The remainder of this paper is organized as follows: Section II reviews related work. Section III details the BLAST system architecture, including the blockchain and agent layers. Section IV presents the auction mechanisms and game-theoretic analysis. Section V describes the experimental setup and results. Section VI provides the conclusion and future work.

\section{Related Work}

The pervasive under-utilization of the radio frequency spectrum necessitates the transition from static allocation to \gls{dsm} and \gls{dsa} paradigms \cite{zhao2007survey, perera2024survey}. Foundational to DSA is the Cognitive Radio, an intelligent system that adapts its parameters to optimize utility and relies on spectrum sensing to identify Primary Users \cite{mitola1999cognitive, haykin2005cognitive, abognah2025decentralized, garhwal2012survey, zhou2025spectrumfm}. 

To manage DSA in complex, multi-user environments, Deep and Multi-Agent Reinforcement Learning (DRL/MARL) are frequently used to identify optimal access policies \cite{jiang2019multi-agent-dsa, gao2021multi-agent-coop-ss, ukpong2025deep, vangaru2024multiagent, chen2025radiollm, rosen2023rfrl}. However, while effective for numerical optimization, conventional MARL often lacks the high-level reasoning required for complex, market-driven spectrum sharing \cite{lee2024llmempowered, karim2025aiagents}.

Decentralized spectrum management requires a trusted, intermediary-free infrastructure, making blockchain and Smart Contracts (SCs) essential for ensuring transaction integrity and automating trading protocols \cite{perera2024survey, cuellar2024blockchain, karim2025aiagents, zheng2020smartcontract, abognah2022distributed}. Blockchain secures critical resource allocation mechanisms, such as Stackelberg games, by providing immutable transaction records \cite{femenias2025stackelberg, wang2025blockchainenabled}. For enterprise-grade privacy and high transaction throughput, permissioned frameworks like Hyperledger Fabric are often adopted, allowing for private data collections that protect sensitive bidding strategies \cite{abognah2022distributed, hyperledgerfabric, fastfabric2019uwaterloo}.

Recently, Large Language Models (LLMs) have introduced intellectual automation to communications. Initially utilized to streamline regulatory tasks via data extraction, Retrieval-Augmented Generation (RAG), and Rules as Code \cite{zhou2024large, rutagemwa2024empowering}, LLMs are now being integrated directly into physical layer processing. Frameworks like RadioLLM \cite{chen2025radiollm} and SpectrumFM \cite{zhou2025spectrumfm} fuse traditional signal extraction with LLM contextual reasoning to achieve state-of-the-art performance in radio classification and spectrum sensing tasks, moving toward Artificial General Intelligence (AGI)-enhanced wireless systems \cite{javaid2025agi, AliZeeshan2026}. 

Concurrently, the deployment of autonomous LLM Agents capable of planning and reasoning has enabled decentralized multi-agent systems for automated spectrum negotiation \cite{karim2025aiagents, hua2024gametheoretic, Lore2024}. Evaluating these agents requires robust game-theoretic frameworks \cite{lee2024llmempowered, jia2024strategic}. However, unlike perfectly rational actors in classical game theory, LLM agents exhibit bounded rationality and frequently deviate from optimal strategies like the Nash Equilibrium \cite{hua2024gametheoretic, herr2024llm, Lore2024}. To mitigate this, researchers impose game-theoretic workflows and structured reasoning onto the agents prior to action selection \cite{hua2024gametheoretic, li2024autoflow}.

Building upon these foundations, our work integrates strategic LLM agents with auction-based spectrum trading mechanisms \cite{hossain2012auctionbased, huang2006auctionbased} to optimize system performance and counteract bounded-rationality limitations. Specifically, we implement a Second-Price Sealed-Bid (Vickrey) Auction \cite{vickrey1961counterspeculation} via a smart contract. This provides a decentralized, provably truthful platform where the dominant strategy is bidding true private valuations, thus ensuring efficient spectrum reallocation, maximizing social welfare, and fostering autonomous market de-concentration.

\section{Problem Formulation and Game Theoretic Analysis}

To facilitate autonomous spectrum sharing, we model the network interaction as a non-cooperative game among wireless operators. We first formalize the general spectrum sharing problem and then derive the optimal strategies for the specific auction mechanisms implemented in the BLAST framework.

\subsection{System Model and Preliminaries}

We represent spectrum access rights as unique, time-limited digital assets referred to as \textit{spectrum tokens} that are traded on a decentralized blockchain. A token $\tau$ represents the right to utilize a specific frequency band for a defined duration within a geographic area. We define a token by the tuple:
\begin{equation}
\tau = \{f_c, B, \Delta t, \mathcal{L}\}
\end{equation}
where $f_c$ is the center frequency, $B$ is the bandwidth, $\Delta t$ is the time-slot duration, and $\mathcal{L}$ is the geographic boundary of operation.

Spectrum trading occurs on the blockchain during predetermined timing ticks and the market is governed by a Smart Contract (SC) that acts as a decentralized automated auctioneer. The SC enforces two critical properties:
\begin{itemize}
    \item \textit{Deterministic Execution:} The auction logic (e.g., clearing rules and winner determination) is hard-coded into the SC, ensuring that the results are verifiable and free from central authority bias.
    \item \textit{Ownership Integrity:} The blockchain provides deterministic finality, preventing "double-spending" of spectrum tokens and ensuring that two operators are not assigned the same frequency band simultaneously.
\end{itemize}

\subsection{Operator Utility and Valuation}
The cognitive radio network consists of a set of wireless operators $\mathcal{N} = \{1, 2, \dots, n\}$, where each operator is represented by an autonomous agent that is tasks with making spectrum trading decisions on behalf of the operator. Each operator's agent $i \in \mathcal{N}$ seeks to acquire a portion of the available spectrum resources to serve their user base. Let $S_i$ denote the amount of spectrum (bandwidth) allocated to operator $i$. 

\begin{figure*}[t]
    \centering
    \includegraphics[width=0.9\textwidth]{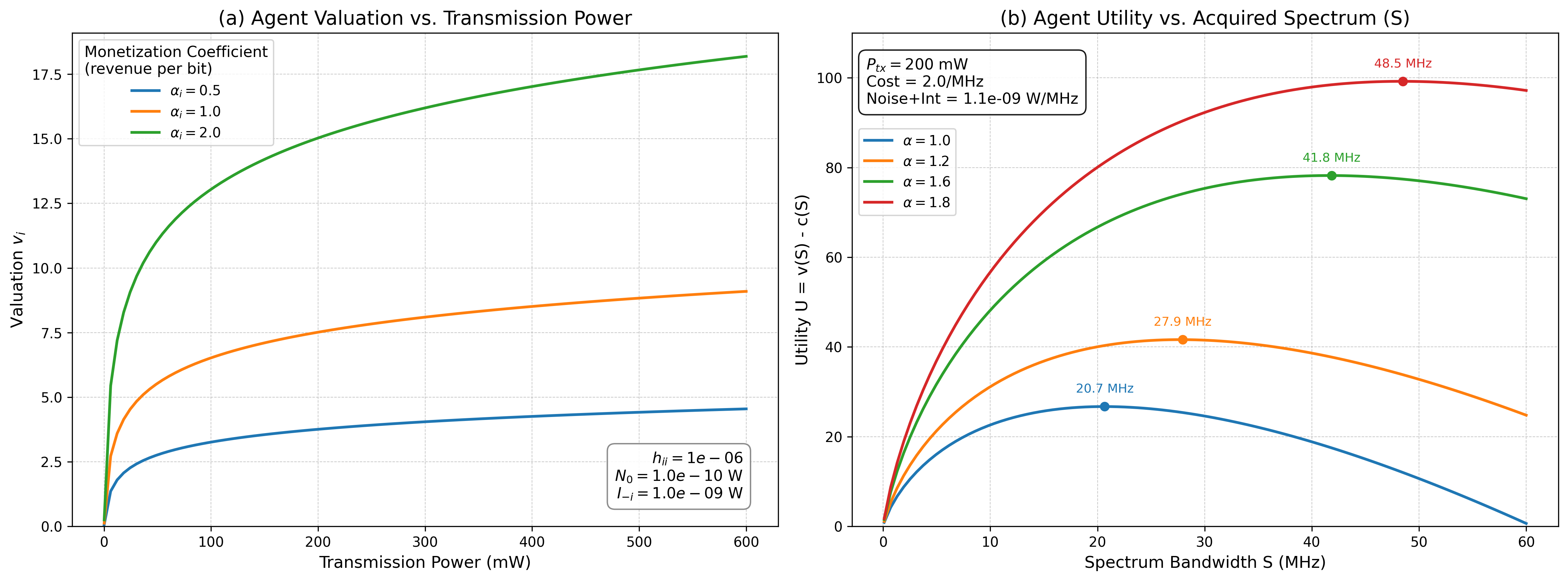}
    \caption{(a) Normalized Valuation vs \textit{SINR} (Linear). (b) Utility vs Spectrum}
    \label{fig:val_and_util_chart}
\end{figure*}

The utility of an operator depends on the quality of service (QoS) they can deliver, which is a function of the acquired spectrum and specific channel state information representing spectral efficiency and \textit{SINR} parameters. We define the valuation function $v_i$ as the monetary equivalent of this utility. A standard form for this utility, based on Shannon capacity, is given by:
\begin{equation}
    v_i = \alpha_i \cdot S_i \log_2\left(1 + \frac{P_i \cdot h_{ii}}{N_0 + I_{-i}}\right)
\end{equation}
where $\alpha_i$ is a monetization coefficient (revenue per bit), $P_i$ is transmit power, $h_{ii}$ is channel gain, and $I_{-i}$ is interference from other users. The logarithmic term $\log_2\left(1 + \frac{P_i \cdot h_{ii}}{N_0 + I_{-i}}\right)$ represents the spectral efficiency (in bits per second per Hertz ($bps/Hz$) based on the \textit{SINR} (Signal-to-Interference-plus-Noise Ratio) which is represented by the term ($\frac{P_i \cdot h_{ii}}{N_0 + I_{-i}}$). Incorporating the \textit{SINR} information in this formulation captures the physical environment characteristics including the diminishing marginal returns in relation to power investment. Figure~\ref{fig:val_and_util_chart}a illustrates this effect and demonstrates that the operator's monetization coefficient $\alpha_i$, which is set by economical and business factors, plays far more significant role in determining the overall valuation of the acquired spectrum.

Thus the total utility of the operator in relation to the acquired spectrum consists of the valuation of the acquired spectrum $v_i(S_i)$ minus the cost to acquire said spectrum $C(S_i)$ as determined by the market clearing price. The objective of each rational agent $i$ is to choose a bidding strategy $\beta_i$ that maximizes their net payoff function $\Pi_i$:
\begin{equation}
    \max_{\beta_i} \Pi_i = v_i(S_i) - C(S_i)
\end{equation}

Figure~\ref{fig:val_and_util_chart}b illustrates the utility function with selected channel parameters by a certain strategy. Traditional optimization models seek to optimize the physical channel parameters (transmission power, SINR, etc..). However, as we can see from the previous analysis, economical factors such as spectrum utilization factor, cost of spectrum, and the amount of spectrum acquired play a much more significant role in the utility maximization. In addition, in multi-operator scenarios decisions made by one operator affect other operators and the optimization problem become intractable.

\subsection{Mechanism Design}
In order to ensure agents act strategically and reach a global optimal solution to the maximization problem, we utilize mechanism design from game theory to construct auction mechanisms that utilize blockchain features such as smart contracts and trusted transaction history.

\subsubsection{Second-Price Sealed-Bid (Vickrey) Auction}
In this mechanism, the highest bidder wins but pays the second-highest bid. We assert that bidding the true valuation ($b_i = v_i$) is a weakly dominant strategy.
\begin{itemize}
    \item If the agent's valuation is higher than the highest competitor bid (\(v_i > b^*\)): Bidding \(v_i\) wins the item and yields a payoff of \(v_i - b^*\). Overbidding yields the same outcome as bidding the true valuation; underbidding below $b^*$ results in losing the item (payoff 0).
    \item If the agent's valuation is lower than the highest competitor bid (\(v_i < b^*\)): Bidding \(v_i\) results in a loss. Overbidding to win forces the agent to pay \(b^*\), resulting in a negative payoff \(v_i - b^* < 0\).
\end{itemize}

Since deviations never increase the payoff and can decrease it, truthful bidding is the dominant strategy.

\subsubsection{First-Price Sealed-Bid Auction}
In this auction mechanism, the winner pays their own bid ($p_i = b_i$). Truthful bidding yields zero surplus ($v_i - b_i = 0$), necessitating bid shading where agents bid less than their valuation in an effort to secure a profit. Bayesian-Nash Equilibrium (BNE) is a known solution for this mechanism \cite{vickrey1961counterspeculation}.

\textit{Bayesian Nash Equilibrium (BNE):}
Assuming $N$ risk-neutral bidders with independent private valuations drawn from a uniform distribution $v_i \sim U[0, \bar{v}]$, the symmetric Bayesian Nash Equilibrium strategy is derived by maximizing expected surplus:
\begin{equation}
    b(v_i) = E[Y_1 | Y_1 < v_i] 
\end{equation} 
where $Y_1$ is the highest valuation among the $N-1$ other bidders. For the uniform distribution, this yields the closed-form strategy:
\begin{equation}
    b^*(v_i) = \left( \frac{N-1}{N} \right) v_i
\end{equation}
This implies that as competition ($N$) increases, bids should approach the true valuation. \cite{vickrey1961counterspeculation}

\textit{Empirical Probability Maximization:}
In our dynamic environment, the distribution of competitor valuations is unknown and might be non-uniform. Therefore, the agent strategy is enhanced by employing an empirical learning based on the bids history on the blockchain.\\
Let $\mathcal{H} = \{w_1, w_2, \dots, w_M\}$ be the set of recent $M$ winning bids $w_i$ observed on the blockchain. The agent constructs an empirical Cumulative Distribution Function (CDF), $\hat{F}(b)$, representing the probability that a bid $b$ will win:
\begin{equation}
    \hat{F}(b) = \frac{1}{M} \sum_{k=1}^{M} \mathbb{I}(w_k < b)
\end{equation}
Where $\mathbb{I}$ is the identity function. The agent then solves the following optimization problem to determine the bid $b^*$:
\begin{equation}
    b^* = \arg\max_{0 \le b < v_i} \left( (v_i - b) \cdot \hat{F}(b) \right)
\end{equation}
This formulation allows the agent to dynamically adjust its shade factor based on the aggressiveness of the current market participants by constantly updating its estimate of the winning probability and the corresponding bidding value.

\subsubsection{Direct Sale (Price Skimming)}
In Direct Sale, the seller sets a "Buy Now" price. To extract maximum surplus, we model this as intertemporal price discrimination (Price Skimming). The seller initiates with a high reserve price to capture surplus from willing high-valuation buyers, then dynamically lowers the price if the item remains unsold.
The initial reserve price $r_0$ is set as a markup over the seller's internal valuation $v_s$ by a markup factor $\beta$:
\begin{equation}
    r_0 = \beta \times v_s
\end{equation}

If the token fails to sell in previous ticks, the price is lowered to traverse the demand curve, subject to a break-even constraint (price cannot fall below the seller's valuation). The reserve price at decision step $t$ is:
\begin{equation}
    r_{t} = \max(v_{s}, r_{t-1} \cdot (1 - \delta))
\end{equation}

Where $\delta$ is the price decay rate. This mechanism effectively segregates the market, capturing high-valuation buyers (urgency buyers) early at high prices, while lowering the price in subsequent epochs to capture lower-valuation buyers, thereby approximating perfect price discrimination over time.

\subsection{System-Level Cooperation and Social Welfare}

While individual agents operate as non-cooperative entities maximizing distinct utility functions, the auction mechanisms function as a coordination layer that enforces system-level cooperation. By aligning individual incentives with resource scarcity, these mechanisms aim to maximize the \textit{Global Social Welfare} (SW). We define the total social welfare as the aggregate of Buyer Profit (Consumer Surplus) and Seller Profit (Producer Surplus). Let $v_i$ be the valuation of the winning agent, $p$ be the clearing price, and $v_s$ be the seller's internal valuation (opportunity cost).
\begin{equation}
    SW = \underbrace{(v_i - p)}_{\text{Buyer Profit}} + \underbrace{(p - v_s)}_{\text{Seller Profit}} = v_i - v_s
\end{equation}

Due to \textit{Transfer Neutrality}, the monetary transfer $p$ cancels out resulting in a \textit{SW} that is a function of the agents' valuations. Maximizing the social welfare then becomes equivalent to \textit{Allocative Efficiency} where the goal is ensuring the spectrum is awarded to the agent with the highest intrinsic valuation $v_i$. To quantify the efficiency cost of decentralization, we utilize the \textit{Price of Anarchy} (PoA). The PoA measures the ratio of the optimal possible welfare ($SW_{opt}$) to the welfare achieved in the worst-case strategic equilibrium ($SW_{eq}$):
\begin{equation}
    PoA = \frac{SW_{opt}}{SW_{eq}} \ge 1
\end{equation}

Where a $PoA$ of 1 indicating optimal equilibrium and less optimal equilibria result in larger values of $PoA$.

\subsubsection{Welfare and PoA in Second-Price Auctions}
The Second-Price auction is theoretically optimal for cooperation. Since the dominant strategy is truthful bidding ($b_i = v_i$), the mechanism guarantees that the agent with the highest valuation $v_{max}$ always wins.
\begin{itemize}
    \item \textit{Welfare:} $SW_{SP} = v_{max} - v_s$.
    \item \textit{Price of Anarchy:} Since the equilibrium outcome is identical to the social optimum, the system exhibits zero efficiency loss.
    \begin{equation}
        PoA_{SP} = \frac{v_{max} - v_s}{v_{max} - v_s} = 1
    \end{equation}
\end{itemize}
This mechanism aligns selfish behavior perfectly with social goals, ensuring no welfare is lost to strategic gaming.

\subsubsection{Welfare and PoA in First-Price Auctions}
In First-Price auctions, cooperation is fragile. Agents shade their bids ($b_i < v_i$) to secure profit. If agents have asymmetric beliefs or risk profiles, a lower-value agent may outbid a higher-value agent who shaded too aggressively.
\begin{itemize}
    \item \textit{Welfare:} $SW_{FP} \le SW_{SP}$. If the optimal agent loses, the realized welfare is based on a sub-optimal valuation $v_{sub}$.
    \item \textit{Price of Anarchy:} The PoA is strictly greater than 1 in asymmetric settings. While the revenue might be higher for the seller, the total system welfare degrades.
    \begin{equation}
        PoA_{FP} = \frac{v_{max} - v_s}{v_{sub} - v_s} > 1
    \end{equation}
\end{itemize}
The increase in PoA represents the "cost" of the agents' strategic uncertainty and bid shading.

\subsubsection{Welfare and PoA in Direct Sale}
In Direct Sales, welfare is threatened by \textit{Deadweight Loss} due to rigid pricing. If the reserve price $r_t$ is set such that $v_s < v_i < r_t$, a mutually beneficial trade is blocked.
\begin{itemize}
    \item \textit{Welfare:} $SW_{DS} = 0$ in the case of a blocked trade, or $v_i - v_s$ for a successful trade.
    \item \textit{Price of Anarchy:} The $PoA$ is highly sensitive to the pricing algorithm. In the case of a blocked trade, the $PoA$ becomes infinite, signaling total market failure:
    \begin{equation}
    PoA_{DS}(t) \approx \frac{v_{max} - v_{s}}{\mathbb{I}(v_{i} \ge r_{t}) \cdot (v_{i} - v_{s})} \rightarrow \infty \text{ as } SW_{eq} \rightarrow 0
    \end{equation}
\end{itemize}

Dynamic pricing (Price Skimming) attempts to mitigate this inefficiency by lowering the PoA toward the ideal value of 1 over repeated time slots. By dynamically adjusting the reserve price, the mechanism ensures that the system eventually clears the market and avoids the infinite PoA associated with total market failure, though the resulting latency introduces a degree of temporal inefficiency.

In this work, we incorporate the above theoretical analysis in an implementation of \textit{Blockchain-based LLM-powered Agentic Spectrum Trading} system \textit{(BLAST)}.

\section{Proposed System Architecture}

\begin{figure}[h]
    \centering
    \includegraphics[width=\columnwidth]{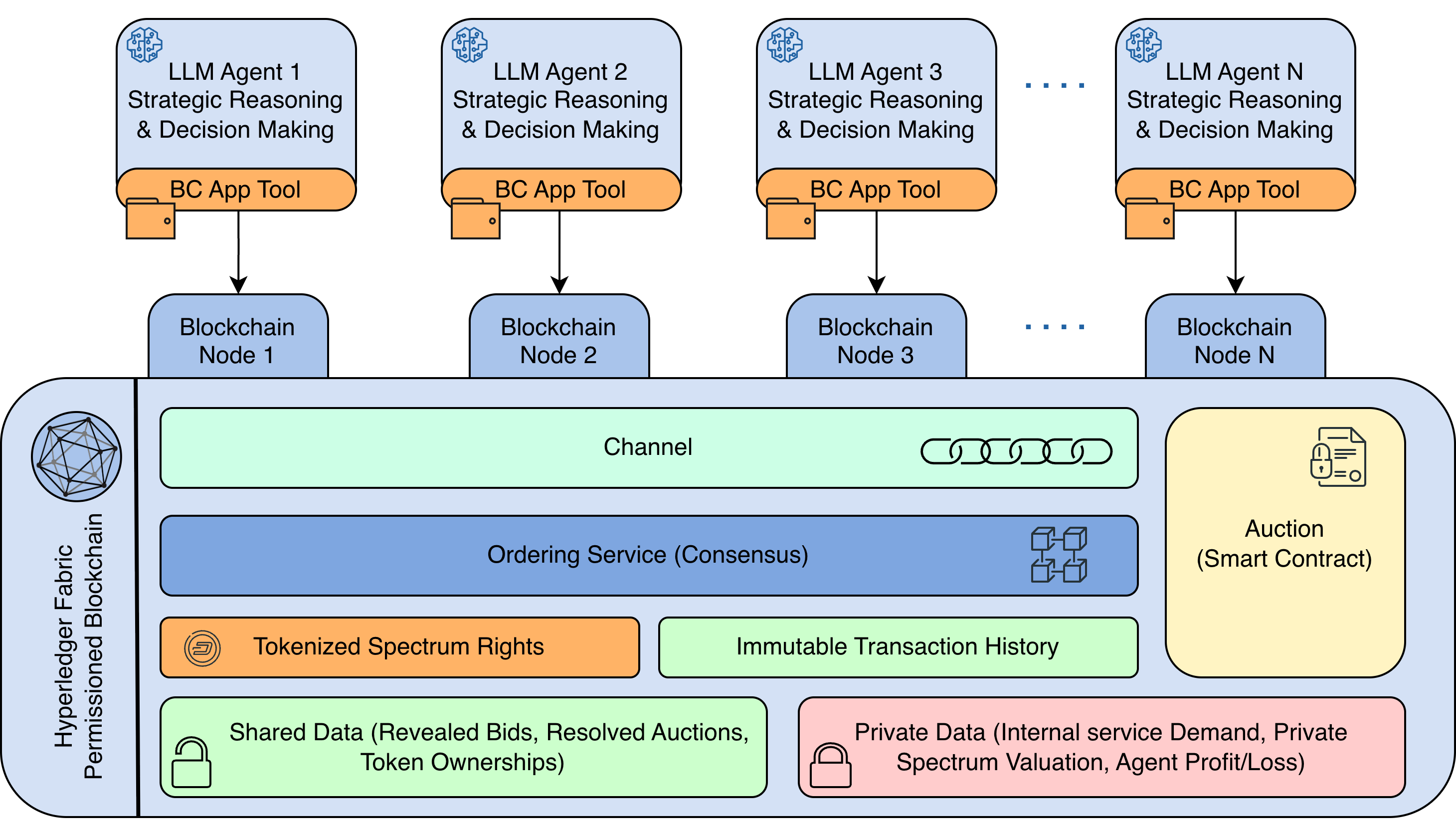}
    \caption{BLAST System Model}
    \label{fig:system_model}
\end{figure}

The proposed BLAST framework (Figure \ref{fig:system_model}) is designed to address the current gaps in decentralized spectrum management by combining LLM Agents reasoning and decision making capabilities with a tokenized spectrum trading marketplace on a blockchain platform that guarantees transparency, trust, and incentivizes users to share their spectrum via game-theoretic profit and system welfare maximizing mechanisms. The proposed module comprises from the Blockchain Layer and the Agent Layer.

\subsection{The Blockchain Layer}

The blockchain serves as the immutable "ground truth" for the system, recording spectrum ownership, managing the lifecycle of tokens, and executing auction logic via smart contracts. The blockchain network is comprised of Nodes representing wireless operators participating in the spectrum sharing marketplace in addition to any stakeholders such as governmental regulators and financial institutions financing and monitoring transactions. Hyperledger Fabric \cite{hyperledgerfabric} was chosen as the technology to implement the blockchain of this model due to it's privacy-focused features, finality of transaction consensus, and identity management which are critical features for a highly regulated applications such spectrum management.

\subsubsection{Permissioned Identity (MSP)}
HLF is a permissioned blockchain where only approved users can access the network via a membership Service Provider (MSP) based on Public Key Infrastructure (PKI) and X.509 certificates. This provides the following critical features that cannot be met with anonymized identities on public blockchains such as Ethereum:
\textit{Accountability:} Every node (Primary User, Secondary User, Regulator) is authenticated before joining.
\textit{Audit Trail:} This creates a legally binding audit trail. If a node causes harmful interference, the regulator can immediately identify the legal entity responsible and revoke their access certificate.

\subsubsection{Granular Privacy via Channels}
In spectrum auctions, particularly sealed-bid formats, users (operators) must keep bid values and strategies confidential to prevent market manipulation.
\textit{Channels:} HLF allows for the creation of private sub-networks (Channels) between specific participants (e.g., a Regulator and a specific Bidder). Competitors on the same network cannot access transaction data or the smart contract state of that channel.
\textit{Private Data Collections:} This feature enables specific data points, such as sensitive interference measurements or the unrevealed bid information, to be shared only with the regulator or kept private until the auction is closed without being written to the global ledger.

\subsubsection{Deterministic Finality}
In dynamic spectrum access, immediate and guaranteed access rights are required once a time slot is purchased.
\textit{No Forking:} HLF utilizes deterministic consensus algorithms (typically Raft). Unlike Proof-of-Work (PoW) or Proof-of-Stake (PoS), where blocks can be reorganized (forked), HLF blocks are final once written.
\textit{Interference Prevention:} This eliminates the risk of ``double-spending'' spectrum tokens, ensuring two operators are not accidentally assigned the same frequency band due to a temporary ledger fork.

\subsubsection{High Throughput and Low Latency}
Cognitive radio networks operate in real-time. The transaction confirmation times of public blockchains are often insufficient for dynamic leasing. HLF employs an \textit{Execute-Order-Validate} architecture, executing transactions in parallel before ordering them. This supports the high volume of micro-transactions required when AI agents autonomously negotiate spectrum access.

\subsubsection{Chaincode Flexibility}
Smart contracts in HLF (Chaincode) can be written in general-purpose languages such as Go, Java, or Node.js. This facilitates the integration of complex logic, including:
\textit{Advanced Auction Logic:} Implementing various auctions types and mechanisms directly in Go provides high flexibility in meeting spectrum sharing needs of the wireless operators.
\textit{Advanced Calculation:} Integrating external libraries and flexible programmatic implementation into the chaincode allows for advanced features to be implemented into the chaincode if needed such as validating as calculate Signal-to-Interference-plus-Noise Ratio (\textit{SINR}) limits before validating a transaction for example.

\begin{table*}[htbp]
\caption{Comparison of Hyperledger Fabric vs. Public Blockchains for Spectrum Management}
\label{tab:hlf_vs_public}
\centering
\begin{tabular}{@{}p{0.07\textwidth}p{0.19\textwidth}p{0.19\textwidth}p{0.46\textwidth}@{}}
\toprule
\textit{Feature} & \textit{Hyperledger Fabric} & \textit{Public Blockchain} & \textit{Impact on Spectrum Management} \\ 
                 &                             & \textit{(e.g.Ethereum)} &                                          \\
\midrule
\textit{Access} & Permissioned (Known users) & Permissionless (Anonymous) & HLF prevents unauthorized radio transmission by unvetted actors. \\ \midrule
\textit{Privacy} & High (Channels/Private Data) & Low (Transparent Ledger) & HLF protects proprietary bidding strategies and sensitive data. \\ \midrule
\textit{Consensus} & Pluggable (Raft/BFT) & PoW / PoS & HLF instant finality, No risk of interference due to double-spending. \\ \midrule
\textit{Cost} & No Gas Fees  & Variable Gas Fees & HLF enables high-frequency trading without unpredictable costs. \\ \bottomrule
\end{tabular}
\end{table*}

The above features of HLF were utilized in the implementation to addresses the critical requirement of privacy and transparency when implementing sealed-bid auctions via the smart contract chaincode:
\begin{itemize}
    \item \textit{Private Data Collections:} We utilize HLF's implicit private data collections. When an agent submits a bid, the actual bid value is stored only on the peers of the organizations involved (the bidder and potentially the endorsers), not on the public ledger.
    \item \textit{Commit-Reveal Protocol:} To ensure integrity while maintaining privacy, we implement a commit-reveal scheme.
    \begin{enumerate}
        \item \textit{Commit Phase:} The bidder submits a transaction containing a cryptographic hash of their bid (salt + value). This hash is recorded on the public ledger, committing the bidder to the value without revealing it.
        \item \textit{Reveal Phase:} After the auction closes, bidders submit a reveal transaction containing the salt and value. The smart contract verifies that $Hash(salt + value)$ matches the committed hash.
    \end{enumerate}
    \item \textit{Endorsement Policies:} Dynamic endorsement policies ensure that the auction outcome is validated by the necessary stakeholders before being committed to the ledger.
\end{itemize}

\subsection{The Agent Layer: LLM-Driven Cognitive Radio}

The LLM agents in BLAST are proposed to provide autonomous and strategic intelligence functionalities mimicking human rational thinking and addressing the shortcomings of manual operational. Employing LLM Agents in the spectrum management framework in general opens the door to a wide range of capabilities beyond standard agents relying on numerical optimization or RL Agents utilizing trial and error to learn the best strategy. LLM Agents can combine information from unstructured data such as regulatory documents or market information into their sophisticated reasoning and chain-of-thought where they can predict competitors moves and anticipate market shifts.

\begin{figure}[h]
    \centering
    \includegraphics[width=\columnwidth]{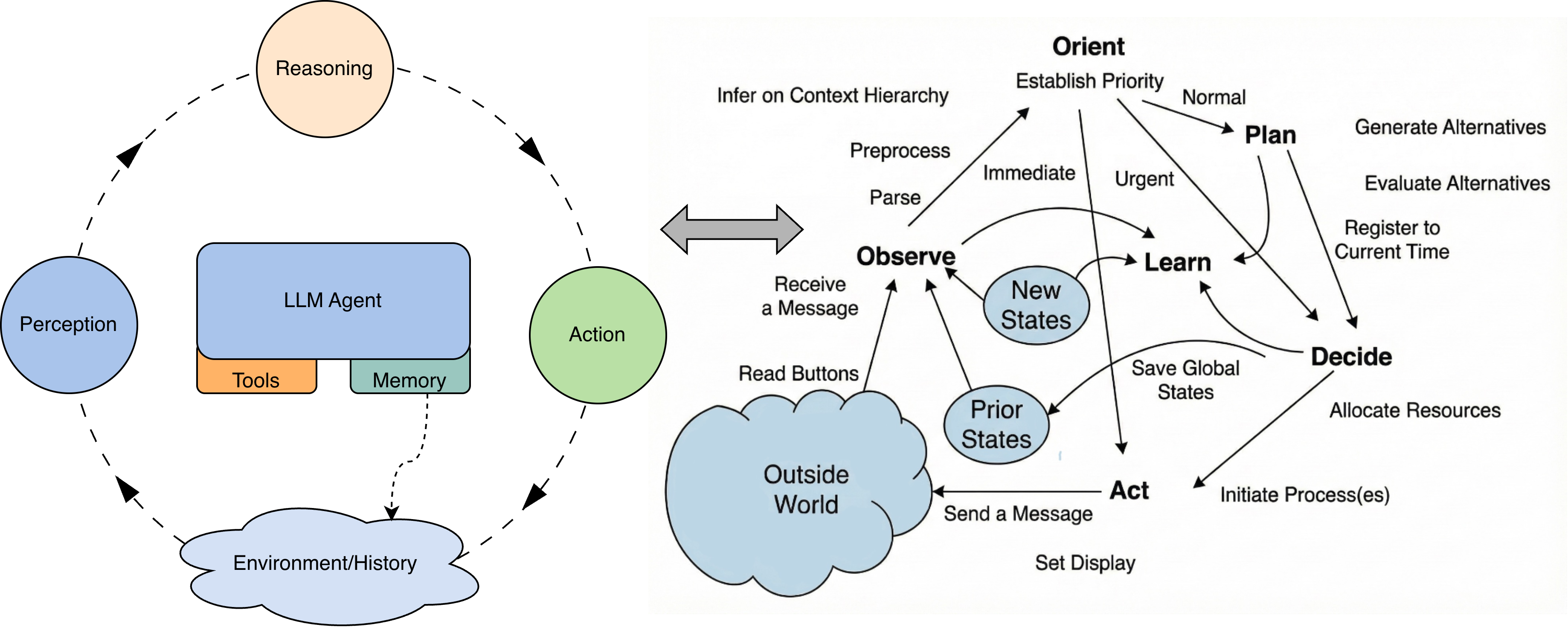}
    \caption{Mapping the BLAST Agent Pipeline to the Cognitive Cycle.}
    \label{fig:agent_cognitive_loop}
\end{figure}

The LLM Agent implementation in this work is structured to execute a sequential pipeline of specialized sub-agents that help maximize the agent's reasoning potential and provide decision explainability and game theoretic reasoning as we will see in the next section. Furthermore, the proposed agent pipeline mirrors the Cognitive Cycle (Figure \ref{fig:agent_cognitive_loop}) originally proposed by Mitola in the original Cognitive Radio paper \cite{mitola1999cognitive} thus enforcing our hypothesis of the validity of this approach.

The proposed agent performed the following steps via sub-agents:

\begin{enumerate}
    \item \textit{Analyst Sub-Agent (Perception \& Orientation)} \\
    This stage ingests raw blockchain data, including transaction history and active auctions. It is tasked with quantifying realized profit per win and classifying the market structure (e.g., "homogeneous" vs. "heterogeneous"). The analyst sub-agent calculates win rates and identifies "demand fading" trends, recording a formal risk assessment before planning begins.
    \item \textit{Planner Sub-Agent (Decision \& Reasoning)} \\
    Acting as the strategic core, this stage utilizes Chain-of-Thought (CoT) reasoning to determines the agent's intent (Buy, Sell, or Idle) based on the spectrum gap and current balance. Crucially, it implements game-theoretic logic adaptable to the auction mechanism:
    \begin{itemize}
        \item \textit{First-Price Auctions:} Estimates the empirical probability of winning to calculate a bid that maximizes expected surplus (Symmetric Nash Equilibrium strategy).
        \item \textit{Second-Price (Vickrey) Auctions:} Enforces a dominant strategy of bidding near true valuation without bid shading.
        \item \textit{Direct Sale (Buy-Now):} Executes price skimming strategies for selling and immediate acquisition for buying if prices are within valuation bounds.
    \end{itemize}
    \item \textit{Action Executor Sub-Agent (Actuation)} \\
    The final stage translates high-level strategy into specific blockchain tool calls. It validates feasibility by ensuring asset ownership for sellers and sufficient budget for bidders before invoking function tooling wrappers such as \textit{start\_auction}, \textit{place\_bid}, or \textit{buy\_now}. This separation of concerns ensures that the LLM's reasoning does not hallucinate invalid transactions.
\end{enumerate}

The implemented agent contains the core AI Agent components:

\begin{enumerate}
    \item Brain: The core of the agent thinking and reasoning capabilities. In this implementation we utilize Google's Gemini models hosted on google cloud servers via the Google Agent Development Kit. This alleviates the need to maintain or develop dedicated LLM models for this application.
    \item Memory: The agent in this model has two types of memory:
    \begin{enumerate}
        \item Internal Memory: This memory is shared by the sub-agents and is used to handover information from one state to another in the agent workflow.
        \item External Memory: Which is the data stored in the blockchain and is accessible to all agents in the system and serves as immutable record of information that is read by each agent at the start of the workflow.
    \end{enumerate}
    \item Persona: Each agent is given a persona that defines its purpose and main objectives and instructs the agent to perform actions that optimize its utility.
    \item Tools: A set of tools that allow the agent to interact with the internal and external memory, handle the workflow transitions, and submit transactions to the blockchain.
\end{enumerate}

\begin{figure}[h]
    \centering
    \includegraphics[width=0.7\columnwidth]{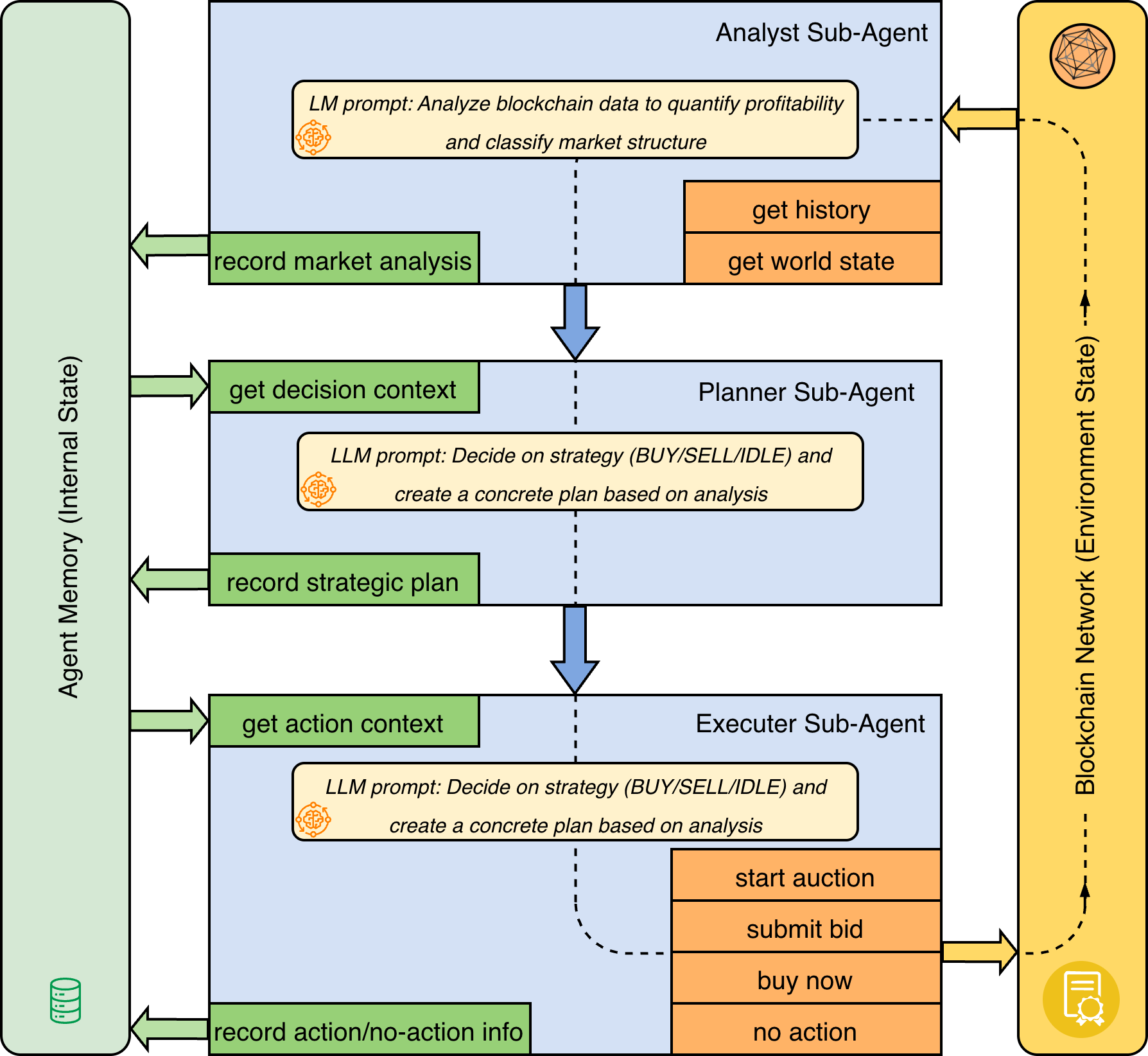}
    \caption{Implemented Cognitive Radio Agent}
    \label{fig:agent_implementation}
\end{figure}

Figure \ref{fig:agent_implementation} illustrates the implemented agent details.

\section{Implementation Details and Experimental Evaluation}

The above proposed model was implemented using Hyperledger Fabric v2 \cite{hyperledgerfabric} and Google Agent Development Kit v1.17 \cite{Google_ADK} calling Gemini 2.5 Flash model with thinking mode enabled in the planner sub-agent. The implementation focused on simulating the behavior of cognitive radio agents that interact with a permissioned blockchain market under two valuation regimes (scenario 1 heterogeneous utilities and Scenario 2 with homogeneous utilities) and three allocation mechanisms (direct-sale listings, first-price auctions, and second-price auctions). Using the execution traces produced by the Google ADK sequential agents, we quantify spectrum utilization, realized profits, and Shapley-value efficiency to measure fairness. In addition, we measure HHI index and Ginin Coefficient to measure market concentration and inequality. The simulation setup and summary of results are presented next.

\subsection{Simulation Setup}
We conducted experiments using the following parameters:
\begin{itemize}
    \item \textit{Agents:} 4 Cognitive Radio Agents (1 Seller, 3 Buyers).
    \item \textit{Tokens:} 25 Spectrum Tokens representing 10MHz blocks in the 3.5GHz band.
    \item \textit{Duration:} 100 simulation ticks.
    \item \textit{Scenarios:} Two scenarios were tested:
    \begin{enumerate}
        \item Scenario 1: Heterogeneous Agents where agents have different spectrum utility 
        \item Scenario 2: Homogeneous Agent where agents have the same spectrum utility
    \end{enumerate}
\end{itemize}

\subsection{Performance Metrics}

The following metrics were collected during the simulation runs and were used to evaluate the performance of the proposed model under different mechanisms.
\begin{enumerate}
    \item \textit{Herfindahl-Hirschman Index (HHI):} Is used as a measure of market concentration and is calculated as $HHI = \sum_{i=1}^{N} s_i^2$, where $s_i$ represents the market share ownership of each agent (the percentage of tokens owned by agent $i$ over the total number of tokens). $HHI < 0.15$ indicates a competitive market; $HHI > 0.25$ indicates high concentration.
    \item \textit{Gini Coefficient:} Measures inequality in profit distribution. The coefficient ranges from 0 to 1, where 0 represents perfect equality (everyone has the same income) and 1 represents maximal inequality (one person has all the income). It is calculated as:
    \[
      \mathrm{Gini} = \frac{2}{n} \cdot \frac{\sum_{k=1}^{n} k\,b_k}{\sum_{k=1}^{n} b_k} - \frac{n+1}{n}
      = \frac{\sum_{i}\sum_{j} |b_i - b_j|}{2 n^{2} \bar b}.
    \]
    Where $b_1 \leq \dots \leq b_n$, are the ordered balances of agents and $\bar{b}$ is the mean balance.
    \item \textit{Social Welfare:} is defined as the sum of the buyer and seller profits in each trade. To calculate this metric for each cleared trade of token \(t\) with capacity \(c_t\) \textit{MHz} between buyer \(b\) and seller \(s\), we track utilities \(u_b\) and \(u_s\) (USD/MHz). Let \(p_t\) denote the realized sell price chosen by the auction rule (respecting the reserve \(r_t\) so that \(p_t \ge r_t)\). Buyer profit:
\[
\pi_b(t) = \max\bigl(0, u_bc_t - p_t \bigr)
\]
In first-price auctions \(p_t\) equals the winning bid; in second-price auctions \(p_t\) is the highest losing bid (or the winner’s bid if no competitor remains); in direct sale \(p_t\) is the posted reserve/ask price. The valuation term \(u_b c_t\) is the buyer’s willingness to pay for the entire block. Seller profit:
\[
\pi_s(t) = p_t - u_sc_t
\]
Here \(u_s c_t\) is the seller’s opportunity cost for keeping the block (aligned with the seller’s reserve). If \(p_t = r_t\), then \(\pi_s(t) = r_t - u_s c_t\) captures the margin over the reserve. With the seller and buyer profits calculated, the total (social) surplus is calculated as the sum:
\[
\pi_b(t) + \pi_s(t) = (u_b - u_s)c_t
\]

    \item \textit{Fairness}: We use Shapley-Values Efficiency to measure how well the implement market mechanism distributes welfare to agents according to their contributions to the coalition. To calculate the coefficient, the seller (\textit{agent-0}) and the three buyers are considered as a cooperative game where every permutation of agent arrivals forms a coalition. For each ordering $S$, we compute the marginal contribution of the arriving agent to the coalition value
\[
v(S) = 
\begin{cases}
0 & \text{if } \textit{seller} \notin S,\\
\left(\sum_{k=1}^{N} \text{c}_{k}\right)\cdot \max_{i\in S} u_i & \text{ow},
\end{cases}
\]

Where $u_i$ is agent $i$'s utility-per-MHz. Averaging the marginal contributions across all permutations yields the Shapley total $\Phi$, i.e., the surplus that a perfectly fair mechanism should distribute. We then compare the cumulative realized buyer\,+\,seller surplus from the blockchain logs to $\Phi$ to obtain a transaction-efficiency ratio $\eta = \text{surplus}/\Phi$.
\end{enumerate}

\section{Results and Discussion}

\subsection{Scenario 1 (Heterogeneous Buyers)}
In this scenario \textit{agent-0} lists all 25 MHz blocks while buyers hold marginal utilities of 10, 15, and 20 USD/MHz. Balances start symmetric at 5,000 USD and the seller alone owns the inventory. Table~\ref{tab:resultsscenario1} compares realized surplus and fairness across auction types.

\begin{table}[t]
  \centering
  \footnotesize
  \setlength{\tabcolsep}{3pt}
  \begin{tabular}{@{}lrrrrrrr@{}}
  \toprule
    Auction & Trades & Avg \$/MHz\$ & Surplus & Shapley & Eff. & Gini & HHI \\
    \midrule
    Direct Sale & 25 & 7.50 & 2775 & 5200 & 53.4\% & 0.104 & 0.347 \\
    First-Price & 33 & 11.06 & 2600 & 5400 & 48.1\% & 0.191 & 0.290 \\
    Second-Price & 25 & 12.80 & 3050 & 4700 & 64.9\% & 0.205 & 0.507 \\
    \bottomrule
  \end{tabular}
  \caption{Scenario-wide market outcomes for Scenario 1 (Heterogeneous Buyers).}
  \label{tab:resultsscenario1}
\end{table}

\begin{figure}[t]
  \centering
  \includegraphics[width=0.7\linewidth]{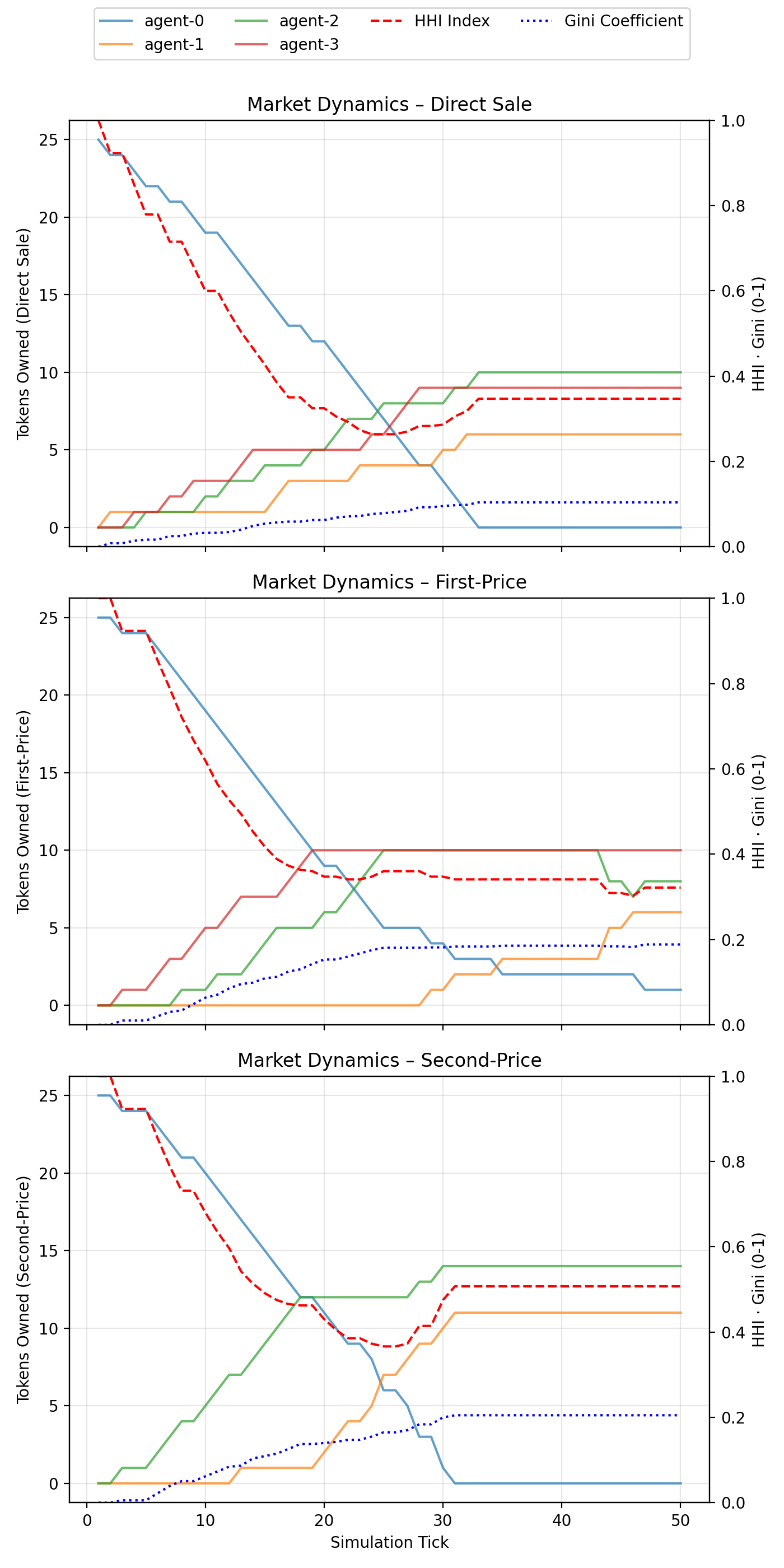}
  \caption{Market dynamics for scenario 1 showing token ownership movements, HHI Index, and Gini Coefficient}
  \label{fig:scenario_1_market_dynamics}
\end{figure}

From the observed behavior in this scenario, we can see that \textit{Direct Sale} cleared 25 trades at an average 7.50\,USD/MHz, delivering 2775\,USD of surplus (53.4\% of the 5200\,USD Shapley benchmark). Balances finished with Gini 0.104 and ownership HHI 0.347, while the average residual spectrum gap stayed at 40.0\,MHz. In the case of \textit{First-Price Auction} the auction cleared 33 trades at an average 11.06\,USD/MHz, delivering 2600\,USD of surplus (48.1\% of the 5400\,USD Shapley benchmark). Balances finished with Gini 0.191 and ownership HHI 0.290, while the average residual spectrum gap stayed at 30.0\,MHz. As for \textit{Second-Price Auction}, the auction cleared 25 trades at an average 12.80\,USD/MHz, delivering 3050\,USD of surplus (64.9\% of the 4700\,USD Shapley benchmark). Balances finished with Gini 0.205 and ownership HHI 0.507, while the average residual spectrum gap stayed at 65.0\,MHz. Figure~\ref{fig:scenario_1_market_dynamics} shows the market dynamics for scenario 1.

\subsection{Scenario 2 (Symmetric High-Value Buyers)}
In this scenario \textit{agent-0} again sells every token, but now each buyer values spectrum at 20 USD/MHz, producing near-identical bids that stress rationing fairness. Table~\ref{tab:resultsscenario2} compares realized surplus and fairness across auction types.

\begin{table}[t]
  \centering
  \footnotesize
  \setlength{\tabcolsep}{3pt}
  \begin{tabular}{@{}lrrrrrrr@{}}
  \toprule
  Auction & Trades & Avg \$/MHz & Surplus & Shapley & Eff. & Gini & HHI \\
  \midrule
    Direct Sale & 25 & 7.50 & 4050 & 5400 & 75.0\% & 0.108 & 0.347 \\
    First-Price & 25 & 16.58 & 3525 & 4700 & 75.0\% & 0.205 & 0.341 \\
    Second-Price & 23 & 20.00 & 4050 & 5700 & 71.1\% & 0.290 & 0.296 \\
    \bottomrule
  \end{tabular}
  \caption{Scenario-wide market outcomes for Scenario 2 (Symmetric High-Value Buyers).}
  \label{tab:resultsscenario2}
\end{table}

\begin{figure}[t]
  \centering
  \includegraphics[width=0.7\linewidth]{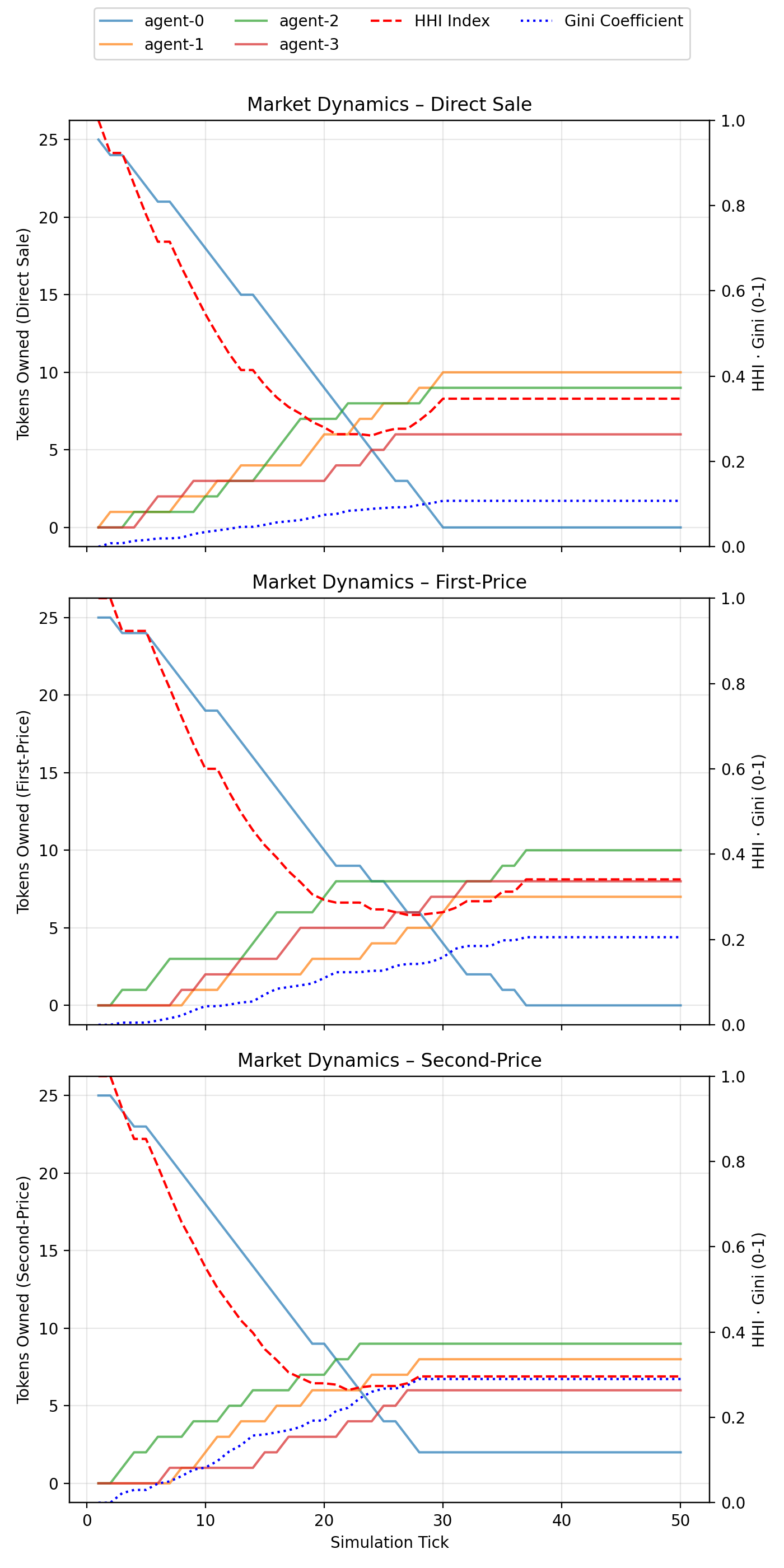}
  \caption{Market dynamics for scenario 2 showing token ownership movements, HHI Index, and Gini Coefficient}
  \label{fig:scenario_2_market_dynamics}
\end{figure}

We can see that \textit{Direct Sale} cleared 25 trades at an average 7.50\,USD/MHz, delivering 4050\,USD of surplus (75.0\% of the 5400\,USD Shapley benchmark). Balances finished with Gini 0.108 and ownership HHI 0.347, while the average residual spectrum gap stayed at 30.0\,MHz. In the case of \textit{First-Price Auction} the auction cleared 25 trades at an average 16.58\,USD/MHz, delivering 3525\,USD of surplus (75.0\% of the 4700\,USD Shapley benchmark). Balances finished with Gini 0.205 and ownership HHI 0.341, while the average residual spectrum gap stayed at 65.0\,MHz. Lastly, the \textit{Second-Price Auction} cleared 23 trades at an average 20.00\,USD/MHz, delivering 4050\,USD of surplus (71.1\% of the 5700\,USD Shapley benchmark). Balances finished with Gini 0.290 and ownership HHI 0.296, while the average residual spectrum gap stayed at 15.0\,MHz, indicating near-balanced utilization. Figure~\ref{fig:scenario_2_market_dynamics} shows the market dynamics for scenario 2.

\subsection{Social Welfare and Efficiency}
\begin{figure}[t]
  \centering
  \includegraphics[width=0.7\linewidth]{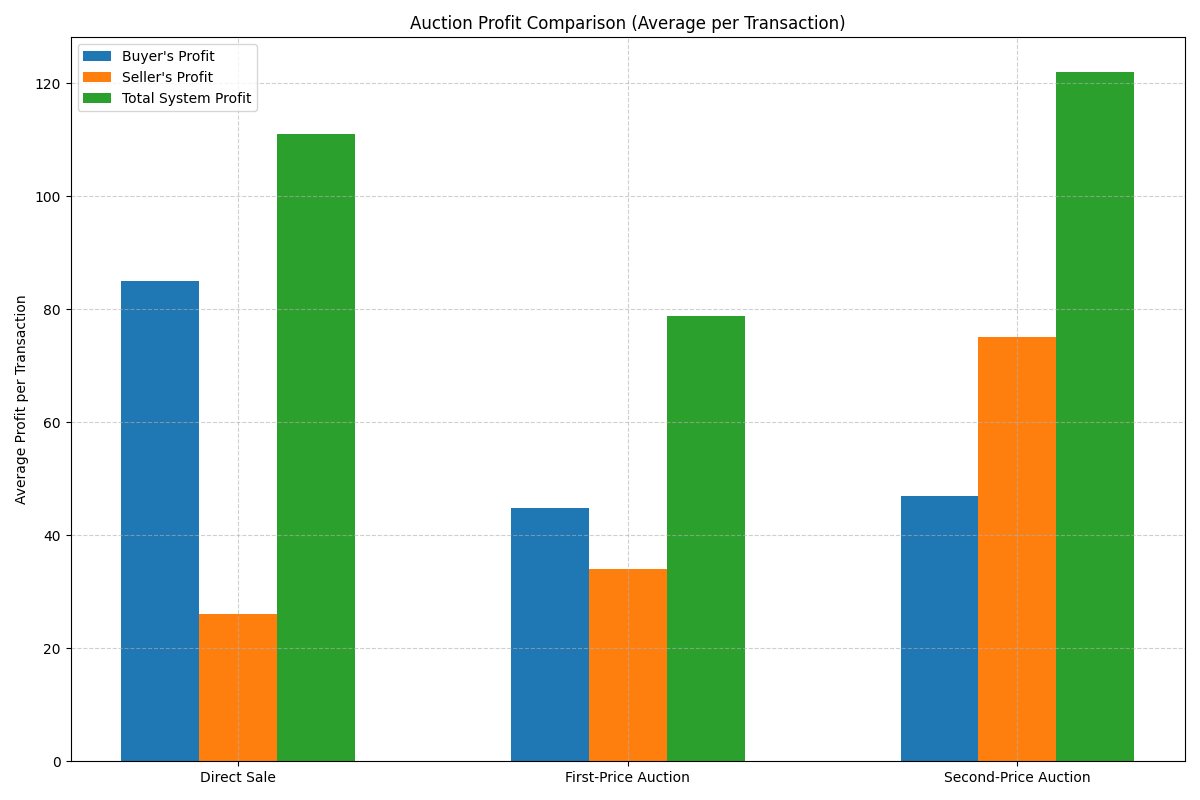}
  \caption{Scenario-1 (Heterogeneous Buyers) Total social welfare highlighting the higher surplus capture of the second-price auction}
  \label{fig:scenario1_welfare}
\end{figure}

In scenario 1, the second-price Vickrey auction consistently achieved the highest social welfare. Because agents bid truthfully, tokens were allocated to the agents with the highest valuations ($v_i$). In the First-Price auction, bid shading occasionally led to inefficient allocations where a risk-averse high-value bidder lost to a risk-seeking lower-value bidder. Figure~\ref{fig:scenario1_welfare} shows the total social welfare for scenario 1.

\begin{figure}[t]
  \centering
  \includegraphics[width=0.7\linewidth]{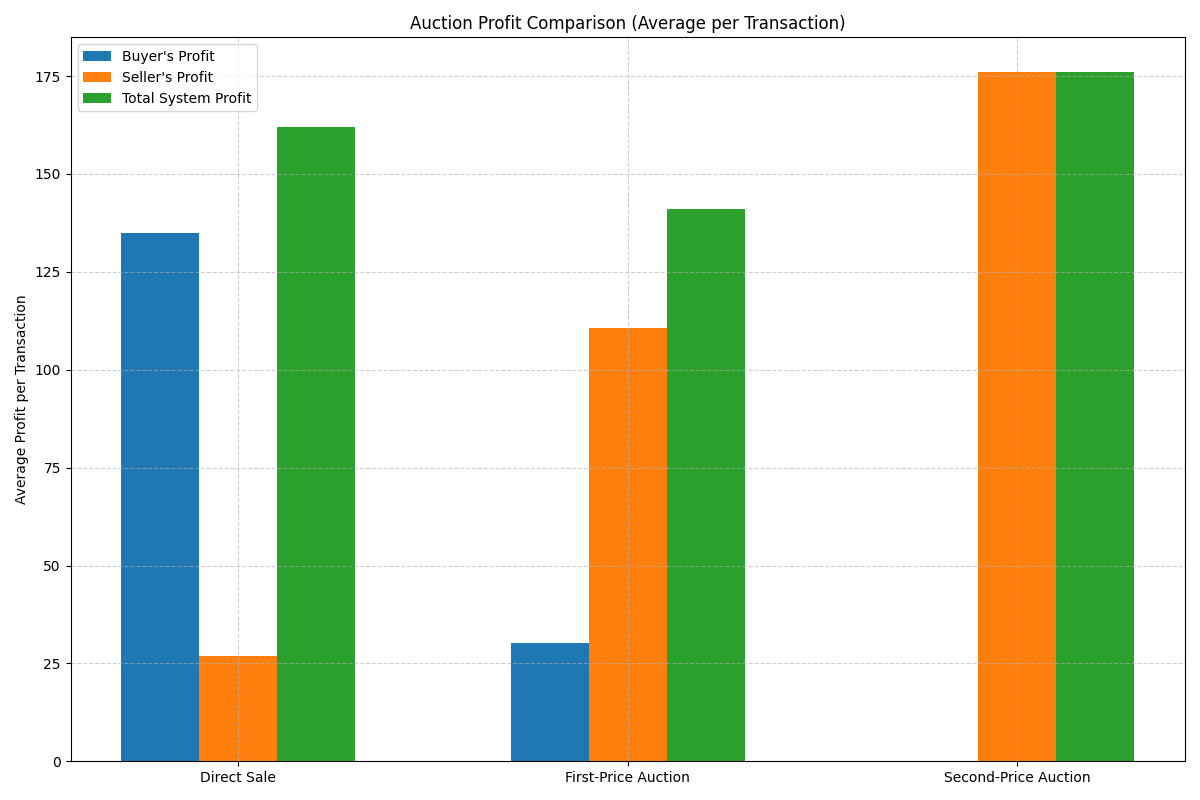}
  \caption{Scenario-2 (Homogeneous Buyers) Total social welfare highlighting the higher surplus capture of the second-price auction}
  \label{fig:scenario2_welfare}
\end{figure}

In scenario 2, due to the equal utility values for each buyer agent, their truthful bidding results in all agents bidding the same values thus eliminating their buying profit in second-price auction as they pay the same amount as their bid. However, even with this outcome, second-price auction still results in the maximum total average profit overall. Results are summarized in Figure~\ref{fig:scenario2_welfare}.

\subsection{Fairness and Shapley-Value Efficiency}

For scenario 1, Direct sale clears inventory quickly and keeps wealth concentration low (Gini $=0.10$), but only captures $53\%$ of the Shapley surplus. First-price auctions under-shoot both fairness and efficiency because aggressive bid shading limits surplus creation even while balances drift toward moderate inequality (Gini $=0.19$). The second-price format achieves the highest efficiency ($65\%$ of the Shapley ideal) by allocating tokens to the highest-value buyers without encouraging shading, albeit at the cost of greater balance dispersion (Gini $=0.21$) and a higher ownership HHI as the most liquid buyer amasses spectrum.

\begin{figure}[t]
  \centering
  \includegraphics[width=0.7\linewidth]{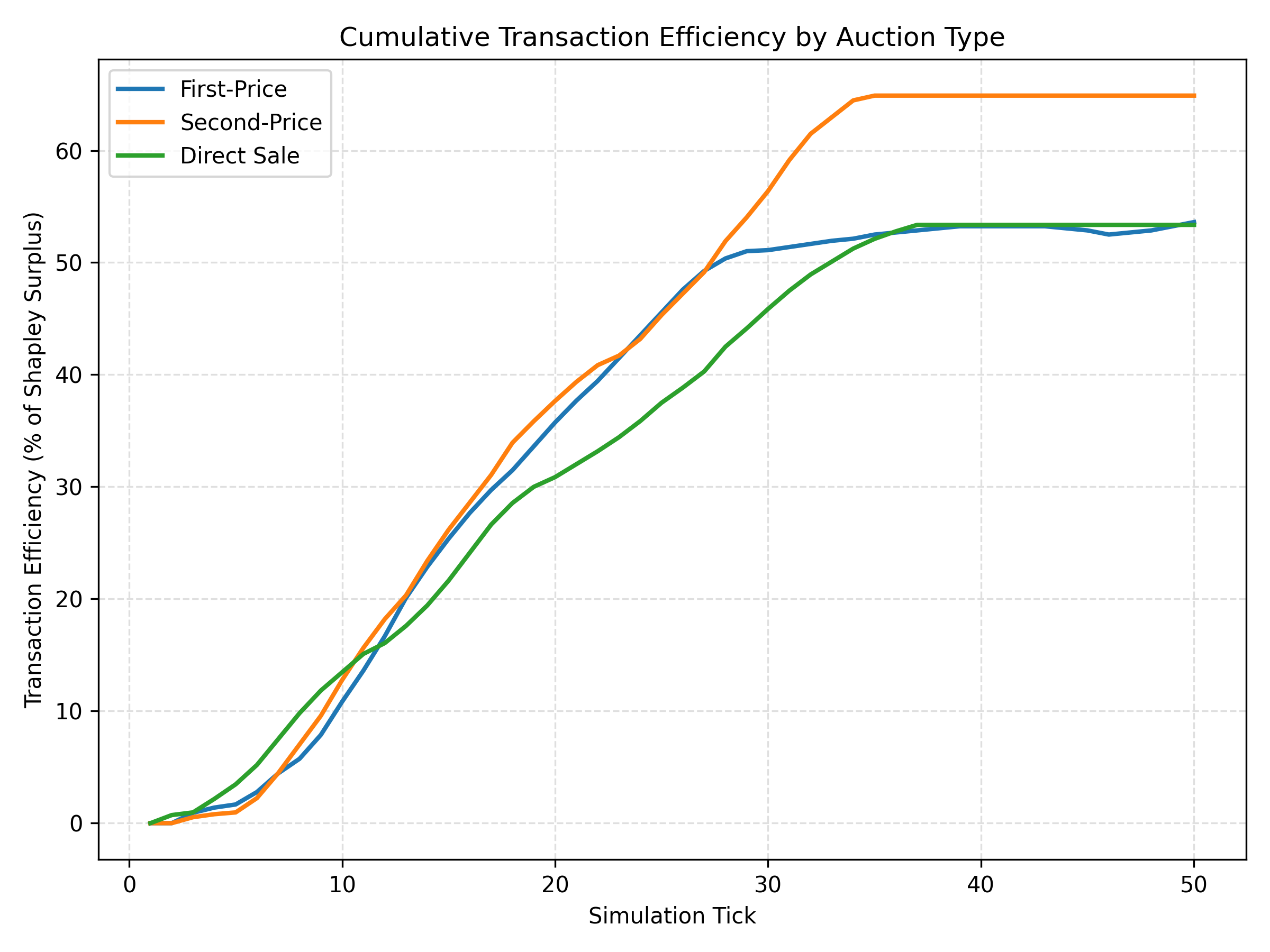}
  \caption{Cumulative transaction-efficiency trajectories for Scenario 1, highlighting the consistently higher surplus capture of the second-price auction.}
  \label{fig:scenario1_efficiency}
\end{figure}

For scenario 2, again, the equal valuations of the buyers results in a slightly lower efficiency for second-price auctions (only 71\% of the shapley ideal compared to 75\% for both direct-sale and first-price auctions. 

\begin{figure}[t]
  \centering
  \includegraphics[width=0.7\linewidth]{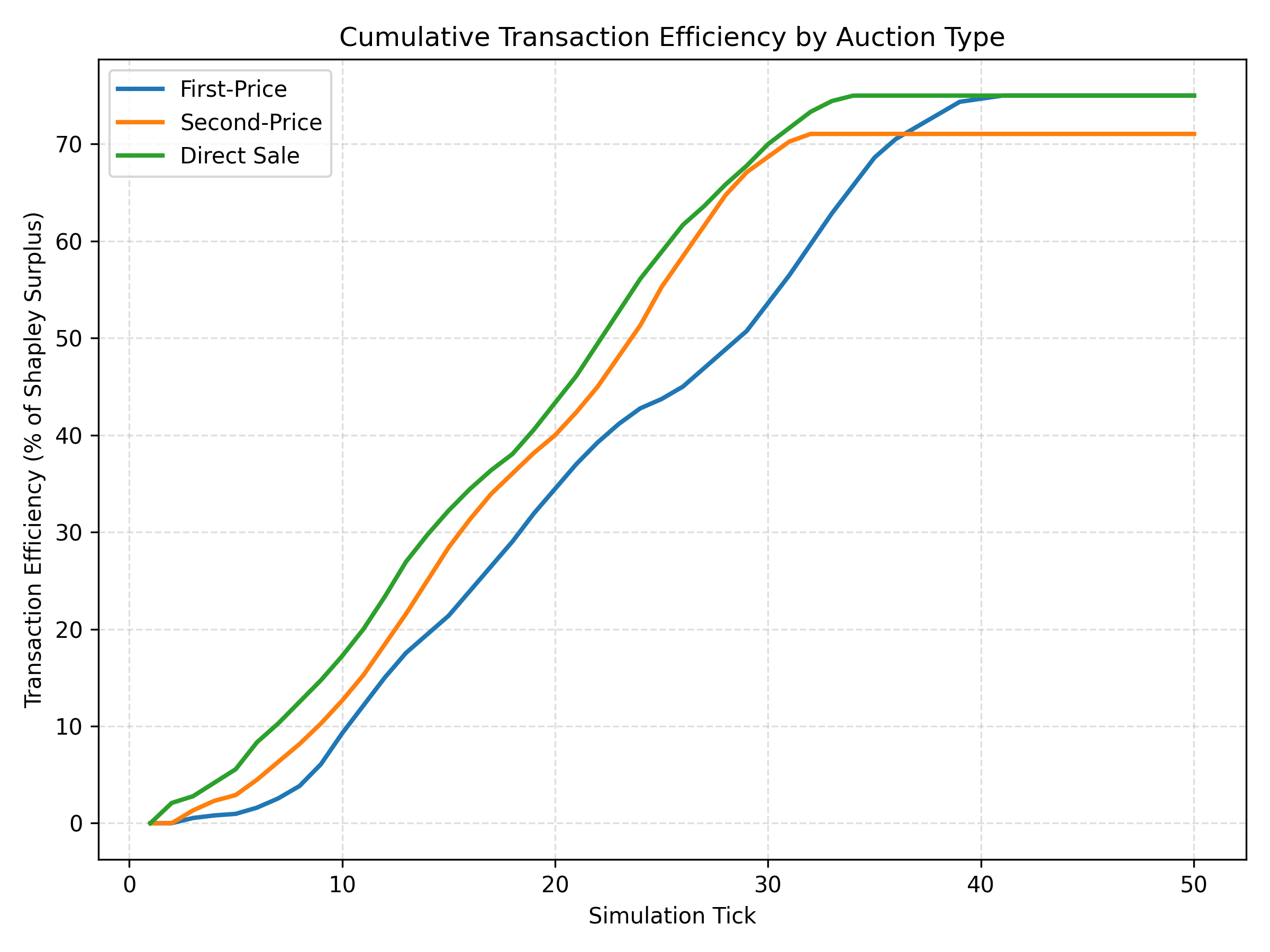}
  \caption{Cumulative transaction-efficiency trajectories for Scenario 2, highlighting the lower surplus capture of the second-price auction due to homogenous market pathology}
  \label{fig:scenario2_efficiency}
\end{figure}

Overall, scenario 1 under-delivers relative to its theoretical welfare frontier: direct sale reaches only 53\% of the \$5.2k benchmark, first-price slips further to 48\% of \$5.4k, and only the second-price format clears a majority of the available value at 65\% of \$4.7k, signaling persistent frictions even in the “best” mechanism. Scenario 2 shows a tighter clustering and higher absolute surplus, with both direct sale and first-price clearing 75\% of their respective Shapley totals (\$5.4k and \$4.7k), while second-price trails slightly at 71\% of a larger \$5.7k opportunity; that consistency suggests the richer demand mix and higher valuations in scenario 2 let each auction recover roughly three-quarters of optimal welfare, whereas scenario 1’s buyers chronically leave surplus on the table regardless of pricing rule.

\subsection{Comparison against Baseline non-LLM Heuristic Agent}

In order to further validate the proposed model, we compared the performance of the LLM-Agents against a non-LLM heuristic agent implementation where buying and selling decision are made based on pre-determined formulations. In this non-LLM scenario, each agent $i$ tracks four parameters: valuation per MHz $u_i$, need profile $N_i(t)$, on-hand spectrum $C_i(t)$, and balance $B_i(t)$. The values of these parameters dictate which action the agent will take in each round (sell/buy/idle).
At tick $t$, the net demand gap is calculated as:
\[
\Delta_i(t)= \max\!\bigl(0,\,N_i(t)-C_i(t)\bigr),
\]
while surplus capacity is:
\[
\textit{surplus} = \max(0,\,C_i(t)-N_i(t))
\]

For any token $k$ with capacity $\kappa_k$, the private valuation is
\[
V_{ik} = u_i \,\kappa_k.
\]

For the \textit{Direct sale mode}, the listing with the largest positive surplus \textit{(valuation - asking price)} is chosen. If the posted price is at or below the agent’s valuation and within its balance, it issues a \textit{buy now} call. When selling, a markup of 15\% is added to the private valuation $V_{ik}$.
In the \textit{First-price auctions} case, buyers shade their bids using a standard symmetric-Nash heuristic: \(( \frac{n_b-1}{n_b}\bigr)V_{ik}\) (where $n_b$ is the number of buyer), but never below half of valuation \(V_{ik}\) and not exceeding their budget. Agents won’t rebid the same token once they have an active offer. Sellers set the reserve price at $1.1V_{ik}$ to reflect the need to cover the winner’s shaded bid.
Finally, in the \textit{Second-price (Vickrey) auctions} bidders simply bid their full valuation (truthful dominant strategy) and sellers list with a light 5\% premium over value since payments clear at the second-highest bid.
In every mode, \textit{the agent will sell only when its owned capacity exceeds current need}, and it lists the first idle token not already on the auction board.
We can already see that multiple simplifying decision had to be made to implement this baseline agent that would otherwise have been made by the LLM model making thus highlighting the benefit of using LLM-Agent model over the heuristic approach.
Furthermore, we ran a simulation using the heuristic agent and compared the resulting parameters for both scenario 1 and scenario 2 described earlier for the three market mechanisms.\\ 

We evaluate the performance of Large Language Model (LLM) driven agents against a heuristic baseline across the two scenarios. In the complex environment of Scenario 1, LLM agents demonstrated aggressive market participation, particularly in auction mechanisms, though efficiency gains were mixed.
\begin{itemize}
    \item \textit{Transaction Volume:} LLM agents exhibited significantly higher activity in auctions, executing \textit{84 total transactions} in First-Price and \textit{54} in Second-Price auctions, compared to the baseline's \textit{25} and \textit{22}, respectively. While some of the LLM transaction are explanatory transactions (result in no sale due to below reserve bidding for example), this indicates a strong propensity for price discovery through repeated trading. Conversely, the baseline executed more Direct Sale transactions (34 vs. 25).
    \item \textit{Efficiency \& Welfare:} The baseline model achieved higher efficiency in Direct Sale (57.1\% vs. 53.4\%) and First-Price auctions (56.6\% vs. 48.1\%). However, the LLM outperformed the baseline in \textit{Second-Price auctions}, achieving \textit{64.9\% efficiency} (\$3,050 welfare) versus the baseline's \textit{55.4\%} (\$2,550). This suggests the LLM's adaptive strategy is particularly effective in truthful mechanisms where aggressive bidding carries less risk.
    \item \textit{Market Concentration:} The LLM agents produced slightly higher inequality (Gini $\approx$ 0.16--0.17) and concentration (HHI $\approx$ 0.39--0.53) in auctions compared to the baseline (Gini $\approx$ 0.13--0.14, HHI $\approx$ 0.40--0.44), reflecting a "winner-takes-all" dynamic in competitive settings.
\end{itemize}

\begin{table}[h]
\centering
\caption{Scenario 1 (Heterogeneous Demand): LLM vs. Baseline Performance}
\label{tab:scenario1_comparison}
\begin{tabular}{l|cccc}
\toprule
\textbf{Mechanism} & \textbf{Tx} & \textbf{Eff (\%)} & \textbf{Gini} & \textbf{Welfare (\$)} \\
\midrule
\multicolumn{5}{l}{\textit{Direct Sale}} \\
LLM & 25 & 53.4 & 0.0878 & 2775 \\
Baseline & 34 & 57.1 & 0.1183 & 2800 \\
\midrule
\multicolumn{5}{l}{\textit{First-Price Auction}} \\
LLM & 84 & 48.1 & 0.1640 & 2600 \\
Baseline & 25 & 56.6 & 0.1367 & 2775 \\
\midrule
\multicolumn{5}{l}{\textit{Second-Price Auction}} \\
LLM & 54 & \textit{64.9} & 0.1737 & \textit{3050} \\
Baseline & 22 & 55.4 & 0.1452 & 2550 \\
\bottomrule
\end{tabular}
\end{table}

In the uniform environment of Scenario 2, the performance gap narrowed, with the baseline offering greater stability.
\begin{itemize}
    \item \textit{Transaction Volume:} Both models achieved identical transaction counts (25) in Direct Sale and First-Price auctions. In Second-Price auctions, the LLM was hyper-active with \textit{103 transactions}, again due to price discovery behavior and competitive bidding due to equal valuation, whereas the baseline remained stable at 25.
    \item \textit{Efficiency \& Welfare:} Efficiency was at parity (\textit{75.0\%}) for Direct Sale and First-Price auctions. In Second-Price auctions, the baseline maintained 75.0\% efficiency (\$4,275), while the LLM dropped slightly to \textit{71.0\%} (\$4,050), indicating some suboptimal allocations due to aggressive bidding.
    \item \textit{Market Concentration:} Gini and HHI metrics were comparable across both models, indicating neither approach led to excessive monopolization in a balanced market.
\end{itemize}

\begin{table}[h]
\centering
\caption{Scenario 2 (Homogeneous Demand): LLM vs. Baseline Performance}
\label{tab:scenario2_comparison}
\begin{tabular}{l|cccc}
\toprule
\textbf{Mechanism} & \textbf{Tx} & \textbf{Eff (\%)} & \textbf{Gini} & \textbf{Welfare (\$)} \\
\midrule
\multicolumn{5}{l}{\textit{Direct Sale}} \\
LLM & 25 & 75.0 & 0.0919 & 4050 \\
Baseline & 25 & 75.0 & 0.1011 & 3975 \\
\midrule
\multicolumn{5}{l}{\textit{First-Price Auction}} \\
LLM & 25 & 75.0 & 0.1650 & 3525 \\
Baseline & 25 & 75.0 & 0.1560 & 3975 \\
\midrule
\multicolumn{5}{l}{\textit{Second-Price Auction}} \\
LLM & 103 & 71.0 & 0.2489 & 4050 \\
Baseline & 25 & \textit{75.0} & 0.2424 & \textit{4275} \\
\bottomrule
\end{tabular}
\end{table}

\begin{figure}[t]
  \centering
  \includegraphics[width=0.7\linewidth]{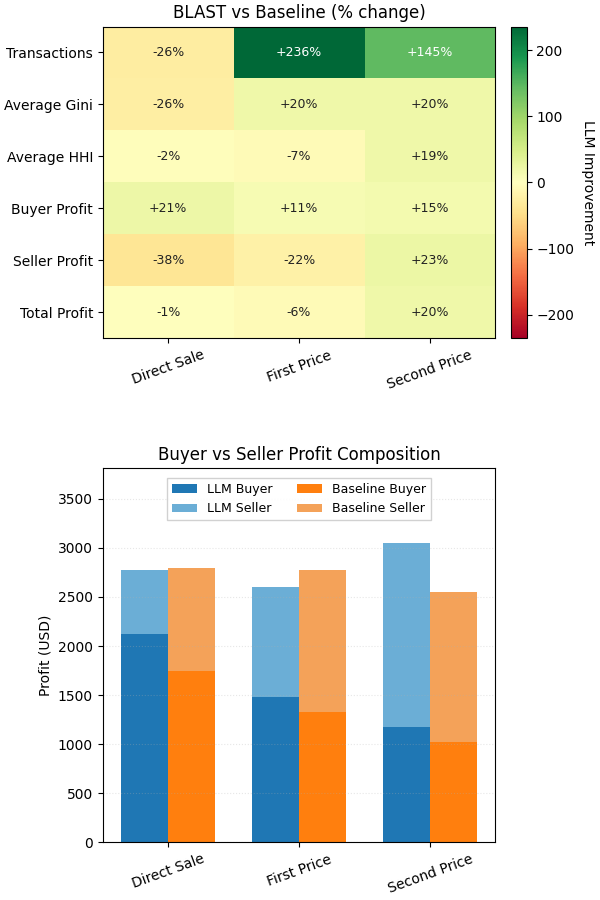}
  \caption{Scenario 1 (Heterogeneous Agents) LLM Agent vs Baseline non-LLM}
  \label{fig:scenario1_llm_vs_baseline}
\end{figure}

\begin{figure}[t]
  \centering
  \includegraphics[width=0.7\linewidth]{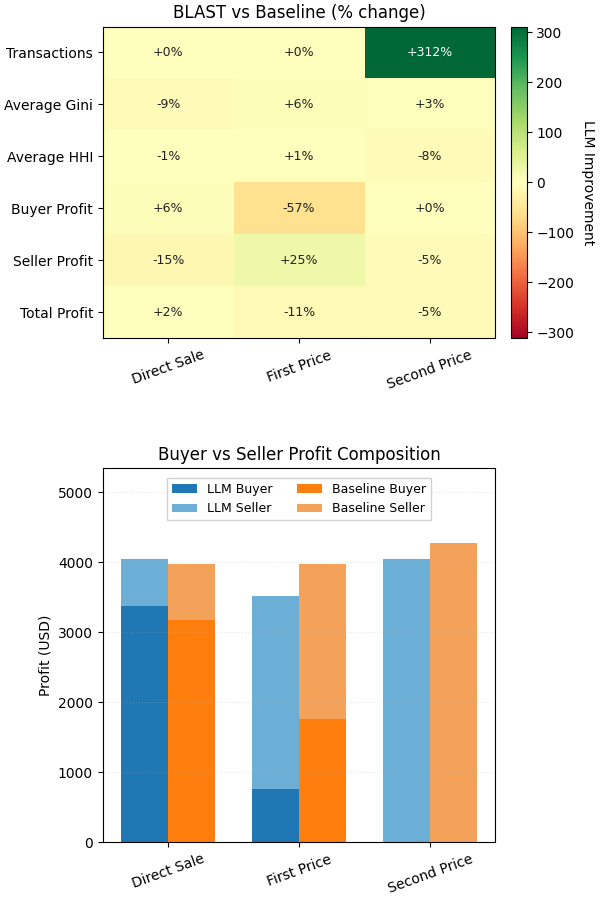}
  \caption{Scenario 2 (Homogeneous Agents) LLM Agent vs Baseline non-LLM}
  \label{fig:scenario2_llm_vs_baseline}
\end{figure}

In summary, the baseline model proves to be a robust standard for stable environments and simple mechanisms, consistently achieving high efficiency with minimal overhead. The LLM model excels in complex, truthful mechanisms (e.g., Scenario 1, Second-Price), where it can outperform the baseline by dynamically discovering pricing and bidding values. Figures \ref{fig:scenario1_llm_vs_baseline} and \ref{fig:scenario2_llm_vs_baseline} visualize these results.

\subsection{Privacy Verification (HLF)}

\begin{figure}[ht!]
  \centering
  \includegraphics[width=0.8\linewidth]{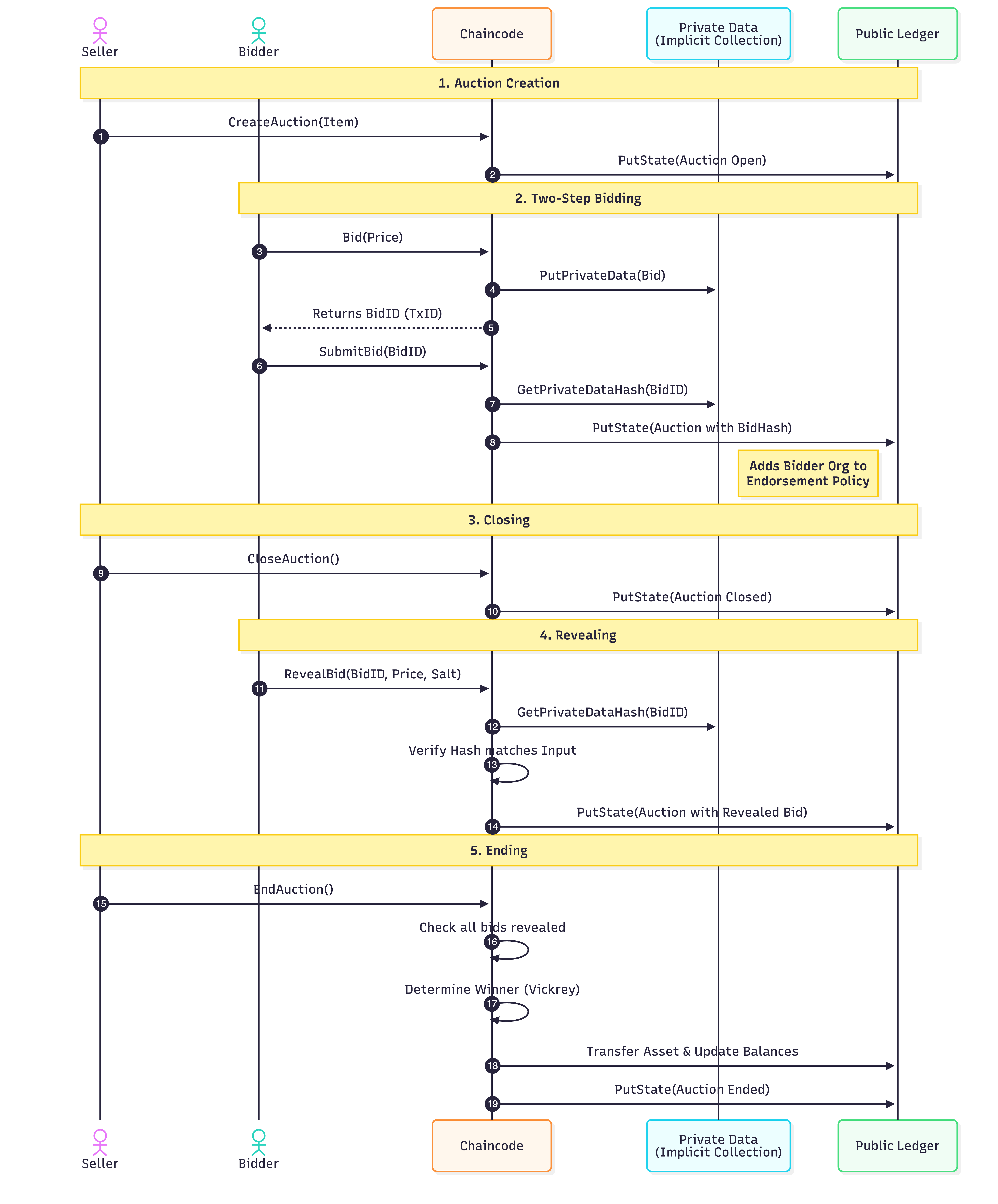}
  \caption{Transaction Diagram of Second-price (Vickery) Sealed-Bid Auction implementation with Hyperledger Fabric}
  \label{fig:hlf-vickery-auction}
\end{figure}

In our Hyperledger Fabric experiments, it was verified that during the bidding phase, the \textit{World State} contained solely the bid hashes. The actual bid values were rendered visible exclusively within the private data collections of the participating peers. This confirms that the system rigorously preserves commercial confidentiality, which constitutes a prerequisite for industry adoption. The implemented sealed-bid second-price auction is shown in Figure \ref{fig:hlf-vickery-auction} illustrating the lifecycle of the auction creation, bidding, closing the auction, revealing the bids, and finally ending the auction and transferring spectrum tokens to the winning bidder and transferring the funds to the seller.

\section{Conclusion and Future Work}

In this paper we presented BLAST, a comprehensive framework for decentralized spectrum trading. By integrating LLM agents with a secure blockchain infrastructure, we demonstrated that the autonomous, strategic allocation of spectrum resources is not only feasible but provides a superior alternative to traditional optimizer-based methods. Our results indicate that LLM agents can effectively participate in various auctions, driving the market towards efficiency and fairness. The proposed model was demonstrated to outperform a non-LLM game-theoretic heuristic model in key metrics. Furthermore, the Hyperledger Fabric implementation rigorously proves that such a system can be deployed with strict privacy guarantees essential for multi-operator environments. Future work will focus on:

\begin{itemize}
\item \textit{LLM Models Benchmark:} The proposed framework can be used to compare the performance of different LLM models and establish a benchmark for model's rational reasoning in the context of dynamic spectrum sharing.
\item \textit{Combinatorial Auctions:} Extending the mechanism to allow bidding on bundles of tokens (e.g., frequency + time + location).
\end{itemize}

This work provides a concrete architectural blueprint for autonomous, efficient, and trustworthy spectrum economies in 6G and beyond, where decentralized intelligence, economic incentives, and regulatory constraints must coexist within a unified execution framework.

\bibliographystyle{IEEEtran}
\bibliography{references}

\end{document}